\documentclass[fleqn,10pt]{wlscirep}
\usepackage[utf8]{inputenc}
\usepackage[T1]{fontenc}

\myexternaldocument{SI}

\title{Engineering optically active defects in hexagonal boron nitride using focused ion beam and water}

\author[1,+,*]{Evgenii Glushkov}
\author[1,+]{Michal Macha}
\author[2]{Esther R{\"a}th}
\author[1]{Vytautas Navikas}
\author[1]{Nathan Ronceray}
\author[3]{Cheol Yeon Cheon}
\author[4]{Ahmed Aqeel}
\author[3,5]{Ahmet Avsar}
\author[6]{Kenji Watanabe}
\author[6]{Takashi Taniguchi}
\author[7]{Ivan Shorubalko}
\author[3]{Andras Kis}
\author[2]{Georg Fantner}
\author[1,*]{Aleksandra Radenovic}
\affil[1]{Laboratory of Nanoscale Biology, Institute of Bioengineering,}
\affil[2]{Laboratory of Nano-Bio Instrumentation, Institute of Bioengineering,}
\affil[3]{Laboratory of Nanoscale Electronics and Structures, Electrical Engineering Institute and Institute of Materials Science,}
\affil[4]{Laboratory of Quantum Nano-Optics, Institute of Physics, Ecole Polytechnique Federale de Lausanne (EPFL), Lausanne, Switzerland.}
\affil[5]{School of Mathematics, Statistics and Physics, Newcastle University, Newcastle upon Tyne, NE1 7RU, United Kingdom.}
\affil[6]{National Institute for Materials Science, Tsukuba, Japan}
\affil[7]{Laboratory for Transport at Nanoscale Interfaces, Empa – Swiss Federal Laboratories for Materials Science and Technology, Dübendorf, Switzerland}

\affil[*]{evgenii.glushkov@epfl.ch, aleksandra.radenovic@epfl.ch}

\affil[+]{these authors contributed equally to this work}

\keywords{hexagonal boron nitride, quantum emitters, optically-active defects}

\begin{abstract}
Hexagonal boron nitride (hBN) has emerged as a promising material platform for nanophotonics and quantum sensing, hosting optically-active defects with exceptional properties such as high brightness and large spectral tuning. However, precise control over deterministic spatial positioning of emitters in hBN remained elusive for a long time, limiting their proper correlative characterization and applications in hybrid devices. Recently, focused ion beam (FIB) systems proved to be useful to engineer several types of spatially-defined emitters with various structural and photophysical properties. Here we systematically explore the physical processes leading to the creation of optically-active defects in hBN using FIB, and find that beam-substrate interaction plays a key role in the formation of defects. These findings are confirmed using transmission electron microscopy that reveals local mechanical deterioration of the hBN layers and local amorphization of ion beam irradiated hBN. Additionally, we show that upon exposure to water, amorphized hBN undergoes a structural and optical transition between two defect types with distinctive emission properties. Moreover, using super-resolution optical microscopy combined with atomic force microscopy, we pinpoint the exact location of emitters within the defect sites, confirming the role of defected edges as primary sources of fluorescent emission. This lays the foundation for FIB-assisted engineering of optically-active defects in hBN with high spatial and spectral control for applications ranging from integrated photonics, to quantum sensing to nanofluidics.
\end{abstract}
\begin{document}

\flushbottom
\maketitle

\thispagestyle{empty}

\section*{Introduction}
Optically-active defects in van-der-Waals (vdW) materials have attracted a lot of attention recently, finding applications in the fields of nanophotonics and nanofluidics\cite{Novoselovaac9439,Atature2018,Radha2016,Esfandiar2017,Fumagalli2018}. In particular, hexagonal boron nitride (hBN) has emerged as a promising material platform, hosting a plethora of optically-addressable defects within its large bandgap ($\approx 6$ eV)\cite{Caldwell2019}. Initially established as bright and stable single-photon sources\cite{Tran2016}, defects in hBN have since been proven to be useful for the integration into hybrid photonic devices\cite{Kim2019},  super-resolution imaging\cite{Kianinia2018,Feng2018,Comtet2019,Glushkov2019,Malein2021,khatri2021stimulated} and studying complex charge dynamics in aqueous environments\cite{Comtet2020,Comtet2021}. Moreover, several recent works have demonstrated spin-dependent emission from defects in hBN\cite{Chejanovsky2021} at room temperature\cite{Gottscholl2020,Kianinia2020,Stern2021}, opening novel avenues for this material platform\cite{Gottscholl2021}.

Defects in hBN are either randomly formed during growth\cite{Caneva2015,Feng2018}, post-growth doping\cite{Koperski2020} or exfoliation from bulk crystals\cite{Choi2016,Yim2020}, or can be intentionally induced in the pristine/unirradiated material using large-area irradiation with ions\cite{Vogl2018,Comtet2019,Fischer2021}, neutrons\cite{Toledo2018}, or using high-temperature annealing\cite{Tran2016}. While these approaches reliably produce quantum emitters in hBN, they result in randomly spatially distributed defects, thus complicating their proper and reliable characterization using correlated microscopy\cite{Hayee2020}, and hindering their applications in integrated devices for nanophotonics\cite{Kim2019} and nanofluidics (e.g. to systematically study the dynamics of charge transfer at liquid-solid interfaces\cite{Comtet2020,Comtet2021}). Recent attempts to generate defects in hBN at precise spatial locations included strain engineering through either transfer/exfoliation\cite{Proscia:18} of hBN flakes or growth of hBN films on patterned substrates\cite{Li2021}, offering scalability, but limited subsequent integration. Other patterning approaches included the use of pulsed lasers\cite{Hou2018,Gao2021}, focused electron\cite{Fournier2020} and ion beams\cite{Ziegler2019,Kianinia2020} to locally damage the hBN lattice and generate emitters with desired properties. 

Focused ion beam (FIB) seems an especially appealing technique for the generation of optically-active defects in vdW materials due to its versatility, high resolution, ease-of-use and scalability.  Notably, FIB has found widespread use for patterning of 2D materials as a resist-free method\cite{Fox2015,Mupparapu2020}, mitigating omnipresent polymer contamination of the patterned material. And emergence of commercially-available plasma FIB (PFIB) machines, employing ions of inert gases, such as argon and xenon, instead of liquid metals, has solved the long-standing problem of sample contamination by gallium ions in the traditional FIB systems\cite{Giannuzzi2011,Zhong2021}. Nevertheless, while deterministic defect generation in hBN using FIB has been recently demonstrated\cite{Ziegler2019,Kianinia2020}, achieving sub-micron spatial accuracy, a clear understanding of the FIB irradiation effects and the formation of defect sites in hBN is lacking. 

In this work, we present a systematic study of these effects, arising from the beam interacting not only with the thin layer of hBN itself, but also with the underlying substrate, performing transmission electron microscopy (TEM) analysis of cross-sections of irradiated hBN flakes. We further show that the FIB-induced defects are strongly influenced by the environment in which the irradiated samples are placed. In particular, we uncover a new mechanism of water-assisted etching of FIB-irradiated hBN defects, leading to  drastic structural and optical transitions in irradiated sites.  Moreover, by utilizing a novel super-resolution microscopy technique correlated with atomic force microscopy (AFM) imaging, we explicitly show the localization of emitters in hBN at the rim of FIB-induced defect sites, in good agreement with a hypothesis that exists in literature\cite{Ziegler2019}.

\begin{figure}[h!]
\centering
\includegraphics[width=0.9\linewidth]{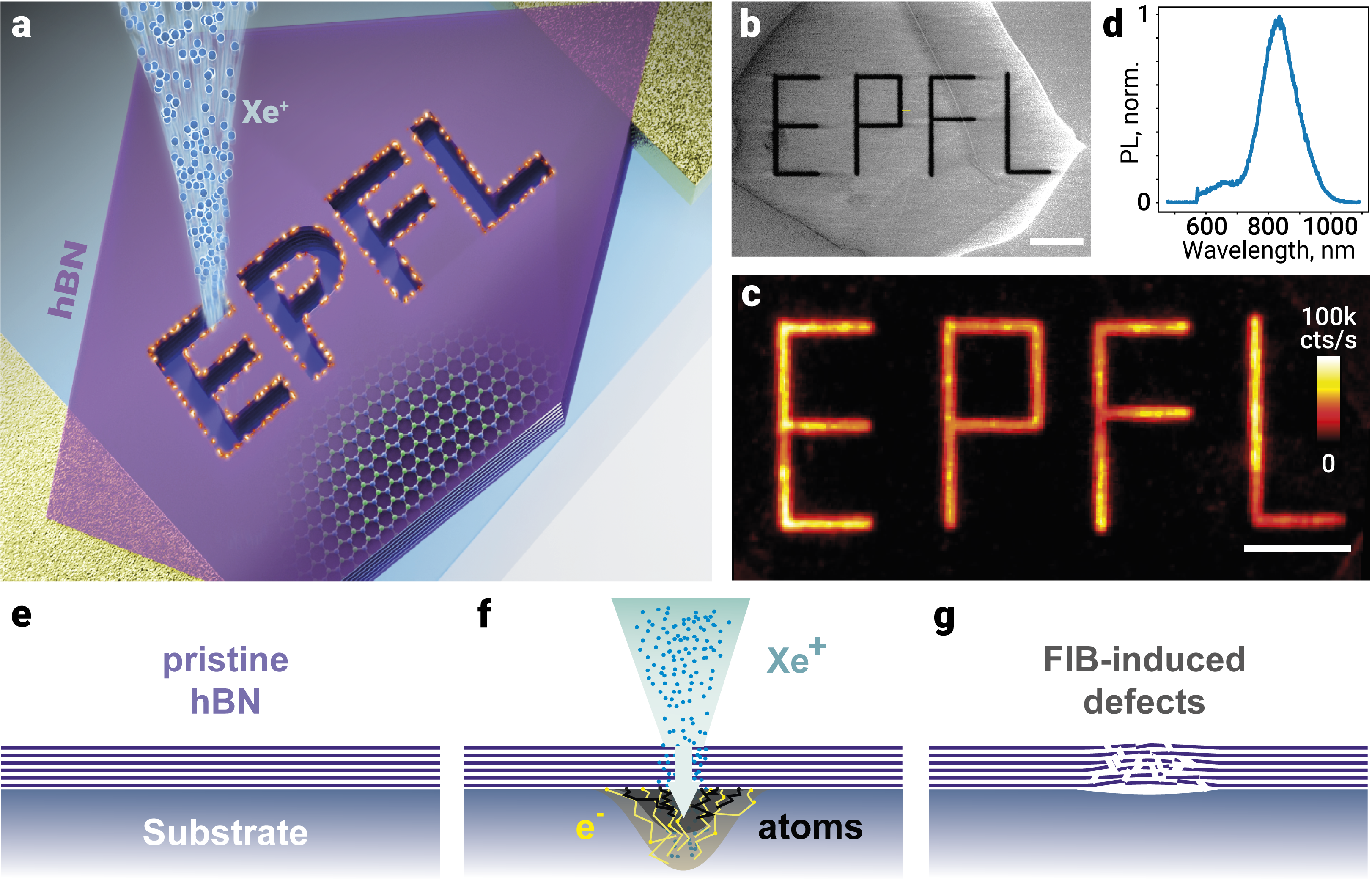}
\caption{Generation of FIB-induced optically active defects in exfoliated hBN flakes. a) Schematic representation of the defect writing process on a pristine/unirradiated hBN flake using Xe FIB. Gold electrodes on the sides reduce charging effects during irradiation. b) A scanning electron microscope (SEM) image of a patterned flake inside the FIB chamber. c) Fluorescence image of the generated pattern of optically-active defects from a confocal microscope. Imaging was done in water. d) Characteristic spectrum of FIB-induced defects, taken from a dried sample in air. e-g) Simplified schematic of the FIB-induced defect generation process: being a result of Xe ions interacting with the substrate, which is partially milled (f), creating a defected region and a void beneath the hBN flake (g).  Scale bars: 10 $\mu$m (a, inset), 5 $\mu$m (e)}
\label{fig:intro}
\end{figure}

\section*{Results and discussion}

We employed the Helios G4 Xe PFIB system to investigate the influence of Xe ion irradiation on pristine hBN flakes, containing very few intrinsic defects\cite{schue2016characterization}. The flakes were prepared on cleaned SiO$_2$ substrates with gold markers and grounding electrodes by mechanical exfoliation from high-quality hBN crystals\cite{taniguchi2007synthesis}, resulting in a typical thickness of hBN flakes of 10-100 nm. Directly after exfoliation, the samples were loaded into the Xe FIB and irradiated with a pre-defined pattern (Fig. 1a). The irradiated flakes were briefly checked in the built-in SEM, clearly demonstrating the FIB-induced changes in the morphology of the sample (Fig. \ref{fig:intro}b). After unloading from the FIB chamber, the samples were inspected on a home-built fluorescence microscope, confirming the creation of optically-active defects at the irradiated sites on hBN flakes (Fig. \ref{fig:intro}c). The spectral characteristics of FIB-patterned defects match the previously reported ones for Xe FIB\cite{Kianinia2020}, with the broad emission peak centered around 830 nm shown in Fig. \ref{fig:intro}d (see Materials and methods for the details). The simplified mechanistic understanding of these changes is shown in Fig. \ref{fig:intro}e-g, where defects are generated as a result of Xe ions interacting with the substrate, creating showers of backscattered electrons and sputtered ions that locally damage the crystalline structure of hBN layers. 

\begin{figure}[ht]
\centering
\includegraphics[width=0.95\linewidth]{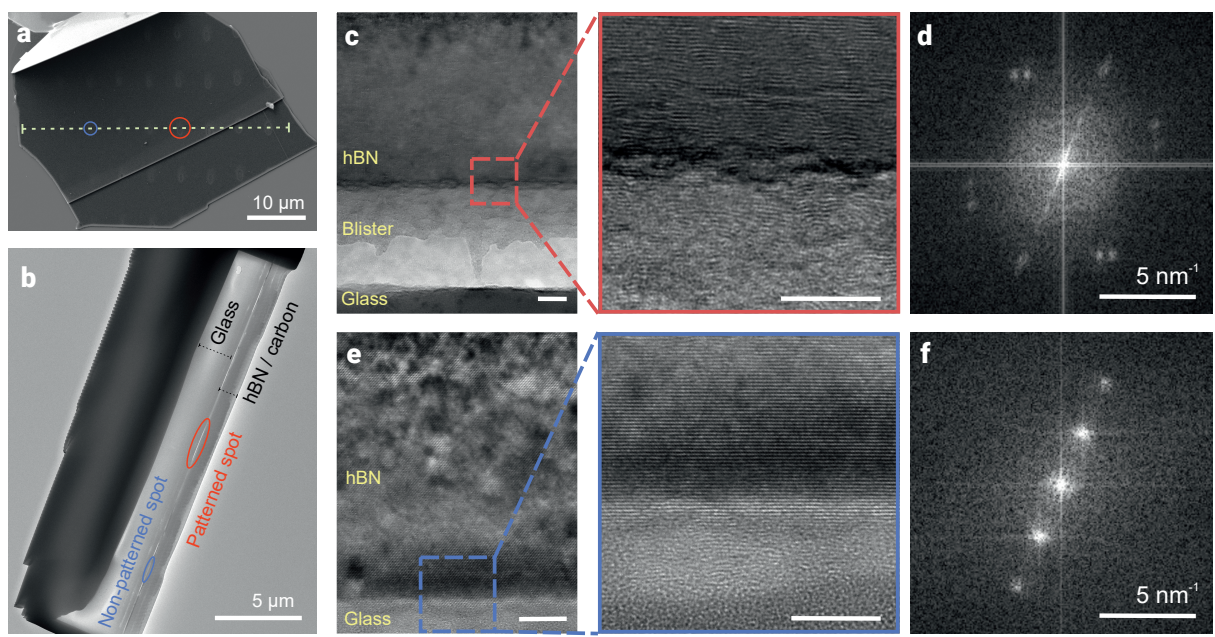}
\caption{Exploring mechanical effects of FIB irradiation using TEM. a) Irradiated hBN flake on a glass substrate, with $3 \mu m$-spaced defect sites (seen as brighter spots on top of a dark hBN surface, marked with the dotted line). b) a lamella of the hBN flake cross-section cut along the dashed line marked in a). A thin layer of carbon was sputtered on top of the flake to prevent charge accumulation. c) A cross-sectional image of the irradiated hBN area and d) corresponding fast Fourier Transform (FFT) compared to e) the pristine/unirradiated area image of the hBN/glass interface and its FFT f) shows an evident mechanical deterioration of hBN layers and the local amorphization of the material.  Scale bars: 10 nm }
\label{fig:lamella}
\end{figure}

To verify the structural changes induced on hBN flakes by the ion beam a detailed high-resolution transmission electron microscopy (TEM) study of the milled area cross-section was performed (Fig. \ref{fig:lamella}). An hBN flake irradiated with typical experimental parameters (see Methods) was cut along the irradiated spots (Fig. \ref{fig:lamella}a) with a Ga FIB system to create a thin lamella (Fig. \ref{fig:lamella}b) suitable for TEM imaging. A high resolution TEM image of the hBN/glass area on the irradiated spot (Fig. \ref{fig:lamella}c) shows visible signs of mechanical damage and perforation caused by the PFIB on the hBN flake layers. Most of the damage is seen at the hBN/glass boundary due to the ion collision cascades, secondary electron emission and substrate particle sputtering occurring in the interaction volume area underneath the patterned spot (schematically marked in Fig. \ref{fig:intro}c). These effects lead to the hBN milling selectively occurring from beneath the flake, at the hBN/glass interface. Consecutively, this leads to the creation of the blister in between the hBN and glass substrate filled with the milled debris comprising of amorphized and recrystalized hBN as well as particles sputtered from within the substrate. The FFT of the irradiated flake area (Fig. \ref{fig:lamella}d) shows signs of substantial amorphization and mechanical damage (smearing of the FFT pattern) of the hBN crystal lattice which can be associated with hBN blistering. The PFIB milling damage is especially visible when compared to unirradiated, pristine hBN/glass interface area (Fig. \ref{fig:lamella}e) and its corresponding, clear FFT (Fig. \ref{fig:lamella}f) image.

\begin{figure}[ht]
\centering
\includegraphics[width=\linewidth]{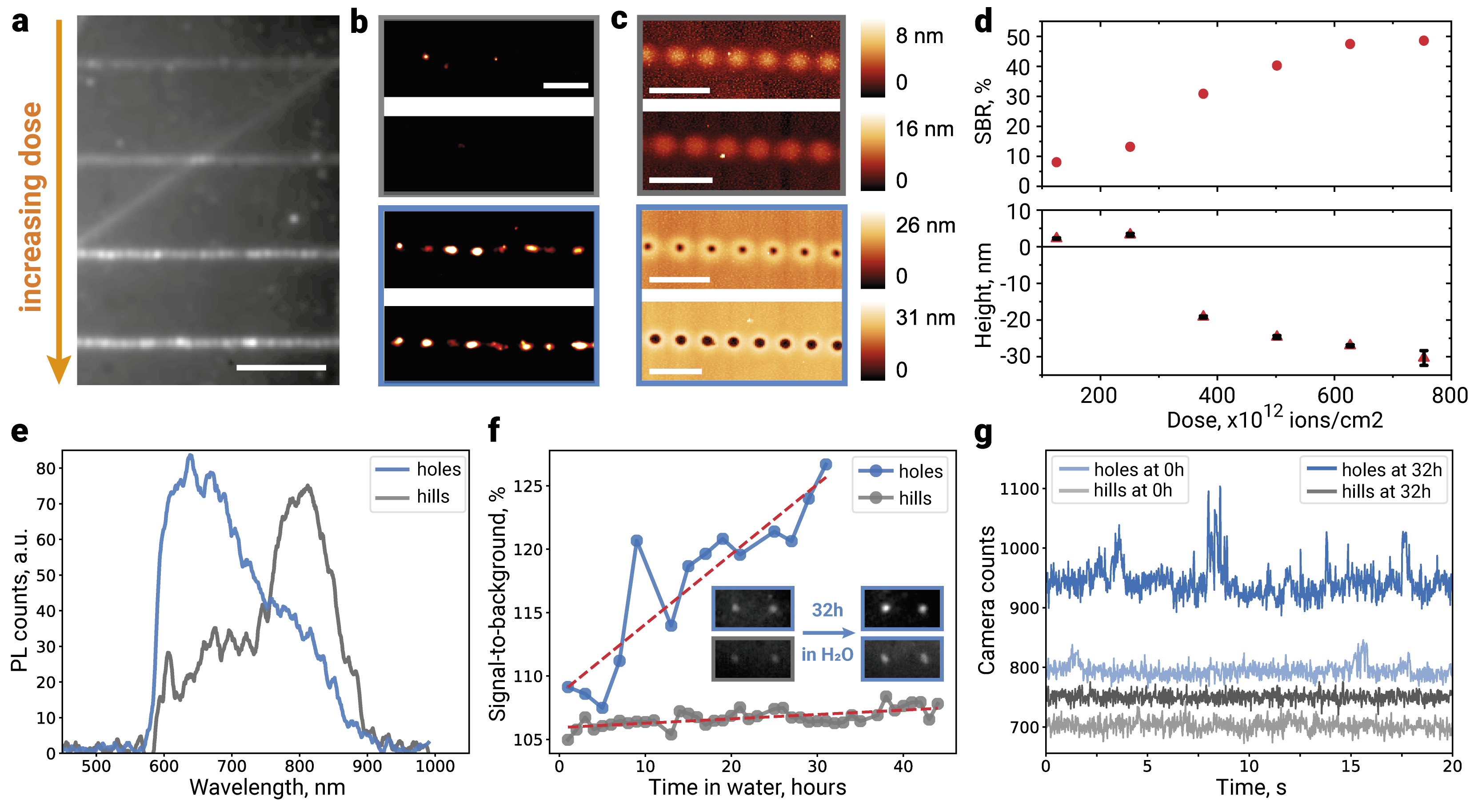}
\caption{Optical and morphological characterization of irradiated defect sites. a) Widefield fluorescence image of an hBN flake with FIB-irradiated horizontal line patterns of a gradually increasing dose, which results in the creation of optically-active defects of increasing intensity. For all measurements in this figure 561 nm laser excitation is used. Imaging is performed in DI water, pH 5.8. b) Zoom-ins from (a), processed using localization microscopy algorithm\cite{ovesny2014thunderstorm}, reveal two different types of emission: dim diffused emitters (type I) which are not localized by the algorithm due to low signal-to-noise (SNR), and bright localized ones (type II) which are clearly visible. c) AFM scans of respective areas. Two types of emitters from optical images can be easily correlated with the difference in morphology, depending on the irradiation dose: hills are formed for lower doses and holes for higher ones. Imaging was done in air, after thoroughly drying the immersed sample with a nitrogen gun. d) Extracted signal intensities (signal-to-background ratio, SBR) and height/depth of the FIB induced defects vs. irradiation dose, showing the dose-dependent hill-to-hole transition. e) Spectra of two defect types, acquired in water. When measuring in air, only the peak around 800 nm is distinguishable (Fig. \ref{fig:intro}d). f) Change in the brightness of two types of emitters after a long immersion in water. g) Comparison of the temporal dynamics of hill vs hole emitters before and after the long immersion. Scale bars: 5 $\mu$m (a), 1 $\mu$m (b,c)}
\label{fig:transitions}
\end{figure}

To explore the optical properties of FIB-induced defects, we have used both a home-built confocal setup and a widefield super-resolution microscope, optimized for single-molecule localization microscopy (SMLM)\cite{Comtet2019}. We started with an hBN flake irradiated with line patterns of varied irradiation dose ($1-8\times 10^{14}$ ions/$cm^2$) to see how it affects fluorescent emission from defects. From the averaged image stack, shown in Fig. \ref{fig:transitions}a, one can see four horizontal lines of varying intensity corresponding to the irradiated pattern of a gradually increasing dose. The zoom-ins into each of these lines in Fig. \ref{fig:transitions}b are processed using an SMLM algorithm\cite{ovesny2014thunderstorm} to precisely localize the isolated emission spots. However, one can see that in the top two, the localization algorithm fails to localize the emitters as their emission intensity is rather dim and diffused resulting in a low signal-to-noise (SNR) and low signal-to-background (SBR) values of $\sim5-10\%$. In contrast to that, the emitters in the bottom two lines are easily localized and are seen as regular bright spots with SBR of $\approx30-50\%$ and $\approx500$ nm pitch distance (Fig. S\ref{fig:si-line-patterns}).

To understand the origin of this emission difference we performed AFM measurements on the same lines (Fig. \ref{fig:transitions}c). One can immediately see that two types of emitters correspond to either hill-type or hole-type structures created by the FIB depending on the irradiation dose. This dose-dependent transition of irradiated hBN areas is shown in Fig. \ref{fig:transitions}d and in more detail in the Supplementary Information (Fig.S\ref{fig:si-line-patterns}). Focusing on the optical properties of the optically-active defects hosted within created hills and holes, we show that their differences go much beyond just emission intensity. In Fig. \ref{fig:transitions}e one can see an appearance of another spectral emission peak around 650 nm, measured from hole defects in water, while hill defects still show predominant emission around 800 nm. While the microscopic origin of this spectral transition is not clear, one can attribute it to the dangling bonds\cite{Turiansky2019} appearing at the edge of the created holes in hBN. The 800 nm emission is also present in the spectrum of holes and is attributed to the remaining damaged material and/or strain around the hole. We further discuss the spatial distribution of these two emission lines in Fig. S\ref{fig:si-spectral-map}.

Another difference concerns the long-term evolution of the measured emission from holes vs. hills when the sample is immersed in an aqueous solution for a long time (tens of hours). In Fig. \ref{fig:transitions}f one can see the steady increase in the signal intensity for holes and an absence of such increase for hills. This graph was obtained by analysing the timelapse images of the irradiated area, fitting the maximal intensity values at each defect site and normalizing it to its local background. An example of such images before and after the timelapse is shown in the inset and the full protocol can be found in Fig. S\ref{fig:si-sbr-calculation}. Finally, the temporal traces from both types of defects in the steady-state, shown in Fig. \ref{fig:transitions}g, demonstrate sharp transitions between different intensity levels (blinking) for the emission originating from holes, but not from hills. This blinking behaviour is further intensified after an extended stay in water which we link to the accumulation of dangling bonds and functional groups in the circumference of hole structures and enhanced interaction with diffusing charges\cite{Comtet2020}. The exact chemical composition of the defect sites is not yet known and remains the topic of future studies. Additional characterization of the optical properties of the created defects is shown in Fig. S\ref{fig:si-optical-characterization}: bleaching, lifetime and saturation curves, as well as optically-detected magnetic resonance (ODMR) at 3.4 GHz closely matching the previously reported one\cite{Kianinia2020} verifying the creation of spin defects in the hBN lattice. Based on our observations and results from literature, one can conclude that the spin-dependent emission comes primarily from the 800 nm peak, for both hole and hill defects, however further studies are needed to clarify this question.

\begin{figure}[ht]
\centering
\includegraphics[width=\linewidth]{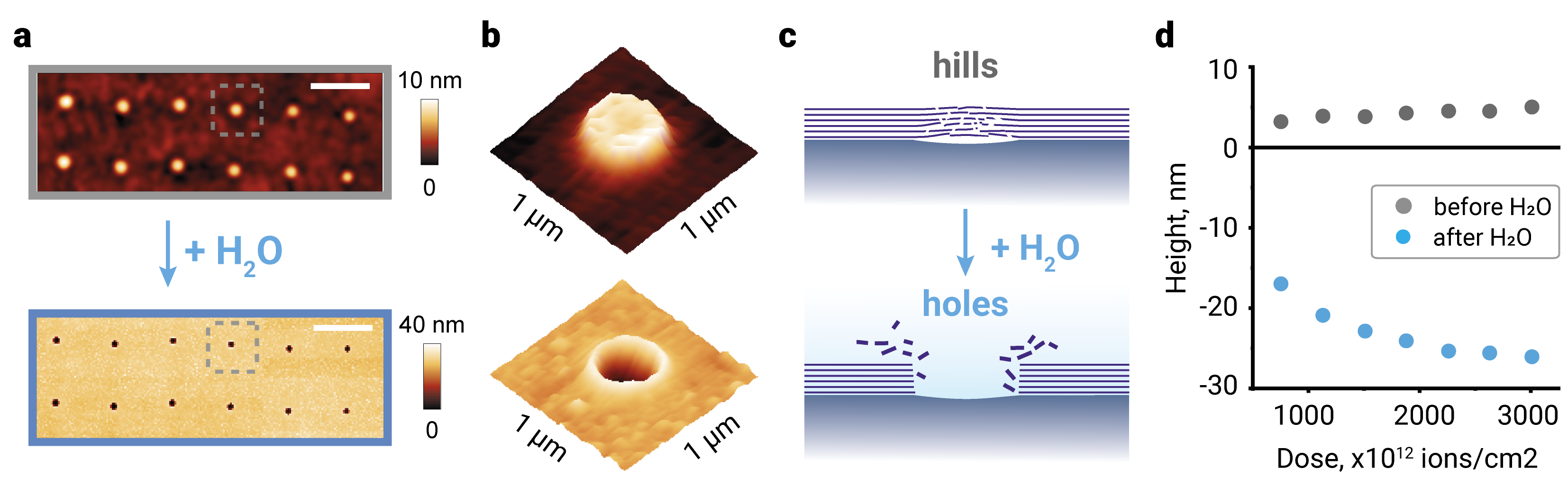}
\caption{Water-assisted etching of irradiated defect sites. a) AFM images of the same area on the irradiated hBN flake before and after immersion in water. Scale bar: 3 $\mu m$ b) Pseudo-3D reconstruction of a single defect site before (top) and after water (bottom). c) Schematic representation of the water-assisted hill-to-hole transition. d) Dependence of the average peak height/depth of irradiated sites on irradiation dose (before and after water). }
\label{fig:water}
\end{figure}

Following the structural-optical transitions between two types of FIB-induced defects, immersed in the imaging medium (DI water), we explored in more detail its influence on the formation of defects. Having obtained an AFM scan of a marked area of an hBN flake just after FIB irradiation, we observed the creation of only hill-type structures, even though the irradiation doses were high enough to create holes (in comparison to Fig. \ref{fig:transitions}). However, after immersing the sample in water and imaging the same area with AFM again we have noticed the immediate creation of holes in the irradiated regions (Fig. \ref{fig:water}a). The observed hill-to-hole transition happened at a timescale of 2 consecutive AFM scans of the same area ($~20$ minutes long), suggesting a fast dissolution process (Fig. S\ref{fig:si-afm-before-after-water}) in comparison to the slow SBR increase of the fluorescent signal in Fig. \ref{fig:transitions}f. The 3D zoom-ins of the selected defect before and after immersion in water are shown in Fig. \ref{fig:water}b and perfectly illustrate the water-mediated hill-to-hole transition. Fig. \ref{fig:water}c shows a simplified schematic of the etching process, where the damaged hBN material at the irradiated sites (seen in Fig. \ref{fig:lamella}) is quickly removed by water. However, once the holes are created their lateral size and depth do not noticeably change both in air and in water over the course of days (Fig. S\ref{fig:si-afm-timelapse-water}). This indicates that only damaged hBN material is removed by water \cite{Streletskii2010}.

The dependence of the created height/depth of the hills/holes, respectively, on the irradiation dose is shown in Fig. \ref{fig:water}d, suggesting a continuous hill-to-hole transition at a lower dose ($\sim3\times10^{14}$ ions/cm$^2$ on Fig. \ref{fig:transitions}d). The slight linear increase in the height of the hills at longer irradiation times probably comes from either more sputtering or a continuous cavity growth underneath hBN, which makes the thin flake bulk up. The saturation behaviour for the depth of the created hills can be explained by water fully etching through the hBN flake and reaching the substrate material. Combining these findings with data from Fig. \ref{fig:transitions}, we conclude that the observed hill-to-hole transition is caused by water and is dose-dependent, i.e. starting from a certain FIB irradiation dose, which mechanically damages the hBN material enough for the water to dissolve it. This claim is supported by the TEM inspection of the irradiated areas in Fig. \ref{fig:lamella} and literature studying the reactivity of mechanically damaged bulk hBN in water\cite{Streletskii2010}. This defect formation process specific of hBN flakes deposited on substrates with thicknesses much larger than the typical beam penetration depth ($<100$ nm)\cite{Kianinia2020}, due to the unique FIB-substrate interaction effects, illustrated in Fig. \ref{fig:intro}e-g and confirmed using TEM in Fig. \ref{fig:lamella}. No hole formation or subsequent structural or optical changes in water were observed for FIB-irradiated hBN flakes which were suspended over holes or supported on thin membranes. An example of such sample and its AFM analysis directly after irradiation is shown in Fig. S\ref{fig:si-tem-grids-afm}, demonstrating the creation of hole defects created by the ion beam in the suspended hBN without any influence of water.

\begin{figure}[ht]
\centering
\includegraphics[width=\linewidth]{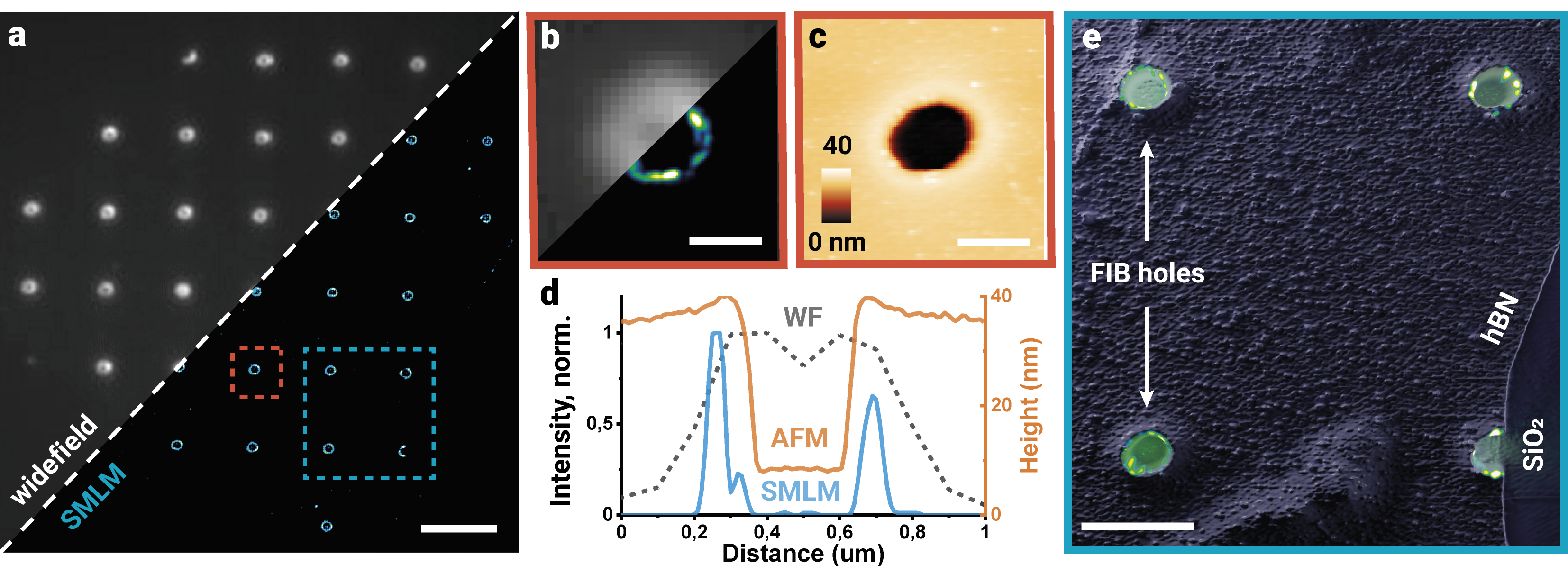}
\caption{Correlated super-resolution and atomic force microscopy of FIB-induced defects in hBN. a) Averaged fluorescence image of an irradiated hBN flake (2D array of isolated defects, 3 $\mu m$ apart), overlayed by the super-resolved image of the same flake (processed using SMLM package DECODE). b) Zoom-in into the dashed red area around the isolated defect site (overlayed widefield/SMLM) c) AFM scan of the same area, perfectly matching fluorescence data. d) Line profiles from zoom-ins in (b) and (c), showing cross-sections of the same defect site in widefield, SMLM and AFM images e) Correlated AFM-SMLM image of four defect sites from a dashed blue area in (a), clearly showing precise spatial localization of optically-active emitters at the rim of the holes, formed by FIB irradiation and activated by water. Scale bars: 3 $\mu m$ (a), 400 $nm$ (b,c), 1 $\mu m$ (e) }
\label{fig:super-res}
\end{figure}

Finally, to better understand the origins of the emission, we performed correlative SMLM and AFM imaging of the hole defects. Briefly, we first acquired stacks of thousands of widefield images of irradiated hBN flakes with isolated defect sites immersed in DI water. The averaged image from one of these stacks is shown in the top-left part of Fig. \ref{fig:super-res}a, while in the bottom-right part we show a reconstructed super-resolved image, processed by a novel deep-learning based localization algorithm DECODE (DEep COntext DEpendent) \cite{Speiser2020}. In some of the bright defect sites in the averaged widefield image one can notice the presence of a darker central region, hinting on the probable emission from the edges of the formed hole as was suggested in literature\cite{Ziegler2019}. Using DECODE to process the signal from FIB-induced defects, we obtain the first direct evidence of such edge-related emission, which is shown in detail in the zoom-in in Fig. \ref{fig:super-res}b. The acquired AFM image of the same area matches well the SMLM data (Fig. \ref{fig:super-res}c). To further verify and visualize the spatial distribution of emitters, we show the overlayed and rendered AFM-DECODE image in Fig. \ref{fig:super-res}d, where one can clearly see the emergence of fluorescent emission from the edge of the hole defects. 

While the hole defects, demonstrated in Fig. \ref{fig:super-res}, are relatively large, smaller sub-100 nm defects were successfully created using the current technique and measured with AFM (see Fig. \ref{fig:water}a and Fig. S\ref{fig:si-afm-hole-size-vs-dose}). However, having dense emitters on the rim of the defect sites hinders their analysis with SMLM-based super-resolution optical microscopy. Further optimization of irradiation parameters and imaging conditions will enable the use of the FIB-created emitters for multiple super-resolution microscopy modalities. Moreover, better spatial resolution and smaller waist of the ion beam, such as in the Helium Ion Microscopes (HIMs), can further decrease the affected defect sites by 1-2 orders of magnitude\cite{Kalhor2014,Buchheim2016,Shorubalko2017}, paving the way for deterministic creation of atomically-small isolated defects with nanometer precision emitting single photons. Further miniaturization of optically-active defects to below 100 nm scale should also enable new studies of nanofluidic phenomena at the liquid-solid interfaces\cite{Comtet2020,Comtet2021} with unprecedented spatial control over the nanoscale charge dynamics.

\section*{Conclusion}

We presented a systematic study of the focused ion beam interactions with supported thin exfoliated hexagonal boron nitride flakes, leading to the creation of optically-active spin defects. We showed that the defect creation is the result of the ion beam interacting not only with the thin hBN flakes, but also with the substrate, on which it is supported. Moreover, both the structural and optical properties of the induced defects are dose-dependent and undergo irreversible transitions in aqueous solutions, representing a new mechanism of water-assisted formation of FIB-irradiated hBN defects. By utilizing super-resolution microscopy correlated with AFM imaging, we were able to explicitly show the localization of emitters at the rim of the created defect site. Our findings lay the foundation for FIB-assisted engineering of optically-active defects in hBN with nanoscale control for nanophotonics, quantum sensing and nanofluidics.

\section*{Methods}

\section*{Sample preparation}
hBN flakes from high quality crystals\cite{taniguchi2007synthesis} were
exfoliated onto glass coverslips (no. 1.5 Micro Coverglass, Electron Microscopy Sciences, 25 mm in diameter), pre-patterned with gold markers for easier navigation and electrode mesh to prevent accumulation of charges. Exfoliation was done either using tape or gel-pack stamps. The flakes for TEM grids (Norcada, NT025C) were polymer-transferred. Si/SiO$_2$ substrates have also been used, with similar patterning results. 

\subsection*{FIB irradiation}
Irradiation and hBN patterning was done on Helios G4 PFIB UXe system with Xenon Plasma FIB column. All experiments were done at 30kV and 100pA Xe beam with varying parameters of dwell time and pitch distance between irradiated spots. By varying the dwell time from $100 \mu s$ to up to 2ms we effectively increase the irradiation dose and pattern outcome in the form of irradiated spot size. Pitch distance is typically set to $3\ \mu m$ for studying isolated defects and between 300 and 500 nm in case of tighter arrays and line patterns. Typical ion fluence/dose ranges from 1.2$\times 10^{14}$ to 2.5$\times 10^{15}$ ions/$cm^2$ for dwell time range of $100\ \mu s$ to 2 ms. Several tens of samples have been routinely created using this technique.

\section*{Optical inspection}
Widefield imaging has been done on a custom widefield fluorescence microscope, described elsewhere\cite{Feng2018, Comtet2019}. Briefly, the emitters are excited using either 488 or 561 nm laser (Monolitic Laser Combiner 400B, Agilent Technologies), which is collimated and focused on the back focal plane of a high-numerical aperture oil-immersion microscope objective (Olympus TIRFM 100X, NA: 1.45). This creates a wide-field illumination of the sample in an area of $\approx 25 \mu m^2$. Fluorescence emission from the sample is collected by the same objective and spectrally separated from the excitation light using dichroic and emission filters (ZT488/561rpc-UF1 and ZET488/561m, Chroma) before being projected on an EMCCD camera (Andor iXon Life 897) with EM gain of 150. An additional spectral path, mounted in parallel to the localization path allows for simultaneous measurements of the emission spectra from individual emitters\cite{Comtet2019}. The sample itself is mounted in a sealed fluidic chamber, which is placed on a piezoelectric scanner (Nano-Drive, MadCityLabs) for fine focus and drift compensation using an IR-based feedback loop which is especially important for long-term measurements. Typical exposure time is 20-50 ms and typical laser power 50-100 mW for the widefield excitation area of  2$\times10^3\ \mu m^2 $, resulting in a power density of $2.5-5$ kW$/cm^2$. A typical acquired image stack contained 2-10 thousand frames.

Confocal imaging was done on two different setups, utilizing either 532 nm or 561 nm excitation lasers and APDs as well as spectrometers to detect the emitted light. The first setup was used to obtain both emission and absorption spectra of defects to make their 2D spectral maps. However, due to the limitations in construction this setup could only be used to image samples in air. In order to perform emission measurements, a diode pumped solid state (DPSS) laser (DJ532-40, ThorLabs) at 532 nm was used. The laser beam was passed through a single mode fiber (P3-460B-FC-1, ThorLabs) to obtain a gaussian beam profile. Afterwards, a narrow bandpass filter (FL05532-1, ThorLabs) was is placed in the beam path to remove any unwanted features from the laser spectrum. A 100X, 0.9 NA objective (MPLFLN, Olympus) focused the beam on the sample and collected the emission. The sample was mounted vertically on a piezoelectric stage (Nano-Drive, MadCityLabs) and a raster scan was performed to obtain the spectral maps. The laser line was removed from the emission spectra by a longpass filter (FELH0550, ThorLabs) and the signal was recorded using a spectrograph (Kymera 193i, Andor) with a CCD (iDus, Andor). For absorption measurement, the sample was illuminated from the back by a calibrated halogen lamp (SLS201L, ThorLabs). The transmitted light was collected, on the opposite side of the sample, by 100X, 0.9 NA objective (MPLFLN, Olympus) and the spectrum was recorded with the CCD. In order to extract the absorption, a measurement was first made on the substrate and used a background signal.

To compensate for this we utilized a second confocal setup in an inverted microscope configuration allowing us to image samples in liquid. Here the emission was split between two fiber-coupled APDs (SPCM-AQRH, Excelitas) in an HBT configuration. One of the APDs could be switched with a fiber-coupled spectrometer (QE Pro, Ocean Optics) to measure the emission spectra in liquid. The setup could also perform lifetime and photon correlation measurements using the PicoHarp TCSPC module (PicoQuant).

\section*{TEM Imaging}
The lamella for cross-section imaging was cut using the Zeiss NVision 40 CrossBeam system. Irradiated flake chosen for cross-section imaging was patterned using default array parameters i.e. 30kV, 100pA Xe FIB. High resolution TEM imaging was performed at Talos L120C G2 using 80kV electron beam.

\section*{AFM imaging and image processing}
For the AFM imaging a customized AFM, consisting of a Dimension Icon AFM head (Bruker) mounted above an optical microscope (Olympus IX83), was used. The position of the hBN flakes with respect to the cantilever was detected with the optical microscope. AFM images in air were recorded at a line rate of $0.5Hz$ with RTESPA-150 cantilevers (Bruker) with a nominal spring constant of $6 Nm^{-1}$ in tapping mode. The cantilever drive frequency and amplitude were determined by automated cantilever tuning. AFM images in fluid were acquired at $0.5 Hz$ line rate using ScanAsyst Fluid cantilevers (Bruker) with a nominal spring constant of $0.7 Nm^{-1}$ in PeakForce quantitative nanomechanical mode (QNM) at an oscillation rate of $1kHz$ and a force setpoint of $3 nN$. The images were processed with a standard scanning probe software (Gwyddion).

\section*{SMLM processing}
Acquired image stacks from the widefield microscope were processed using several SMLM algorithms: ThunderSTORM\cite{ovesny2014thunderstorm}, SOFI\cite{} and DECODE\cite{Speiser2020}, a new deep learning-based localization algorithm. DECODE processing is beneficial in this case as the emitter density around the rim is high (Fig. S\ref{fig:si-smlm-montage}) and standard SMLM algorithms, such as Thunderstorm\cite{ovesny2014thunderstorm}, are failing to properly localize individual emitters, resulting in localization artifacts (Figs. S\ref{fig:si-smlm-montage},\ref{fig:si-wf-ts-sofi}). To utilize DECODE, We trained the neural network provided by the authors using simulated frames with a high density of emitters (5 $\mu m^{-2}$), a realistic intensity distribution (1000 $\pm$ 800 photons/event), and the experimental point spread function calibrated with fluorescent beads.

Another approach to analyse such type of data with dense emitters is to use the super-resolution optical fluctuation imaging (SOFI), which better tolerates the high density of emitters and similarly to DECODE shows the clear existence of the fluorescent rim at the edge of hole defects (Fig. S\ref{fig:si-wf-ts-sofi}). SOFI images were processed using a previously published algorithm\cite{Geissbuehler2012}. Both SOFI and DECODE approaches have allowed to clearly resolve rims of the defect sites, but DECODE led to fewer image artifacts than SOFI.

\section*{Acknowledgements}
The authors would like to acknowledge support from the EPFL Interdisciplinary Center for Electron Microscopy (CIME) and especially Lucy Navratilova for the help with lamella preparation.  E.G. V.N.and A.R. acknowledge the Max-Planck EPFL Center for Molecular Nanoscience and Technology as well as the NCCR Bio-inspired materials. M.M. acknowledges financial support from  Swiss National Science Foundation (SNSF)  Grant Number $200021_192037$.  K.W. and T.T. acknowledge support from the Elemental Strategy Initiative conducted by MEXT, Japan, and CREST (JPMJCR15F3), JST.

\section*{Author contributions}

A.R., E.G. and M.M. conceived and designed the experiments; E.G., A.A. and C.Y.C. prepared the samples and M.M. performed the FIB irradiation; E.R. did the AFM measurements, helped by N.R. and C.Y.C. E.G. performed the optical measurements with help from V.N., A.A. and N.R.; M.M. did the TEM measurements; E.G, V.N. and N.R. analyzed the data; K.W. and T.T. contributed materials; I.S. helped with interpretation of results; E.G. wrote the paper, with inputs from all authors; A.K., G.F. and A.R. supervised the project; All authors discussed the results and commented on the manuscript.

\section*{Data availability}
The data that support the findings of this study are available from the corresponding authors on reasonable request.

\section*{Competing interests}
The authors declare no competing interests.

\section*{Additional information}
Supplementary information is available in the online version of the paper.

\bibliography{main}

\end{document}


\flushbottom
\maketitle
\newpage
\listoffigures

\thispagestyle{empty}

\newpage

\begin{figure}[ht]
\centering
\includegraphics[width=\linewidth]{figures/SI/hBN lamella fig supp1.pdf}
\caption[TEM micrographs of FIB-cut hBN lamella]{TEM micrographs of FIB-cut hBN lamella. Each subsequent image a-f) shows a detailed structure of the irradiated cross-section with increasing magnification and focus on the interior of the patterned spots. }
\label{fig:si-lamella-closeup}
\end{figure}

\begin{figure}[ht]
\centering
\includegraphics[width=\linewidth]{figures/SI/fig-S2.png}
\caption[Analysis of FIB-induced line patterns of structural and optically active defects]{Analysis of FIB-induced line patterns of structural and optically active defects. a) Fluorescence image of an exfoliated hBN flake, with lines of defect sites, induced by the FIB. Irradiation dose increases from the top to the bottom line. In-line pitch: 500 nm. b) Average intensity of each line of defects. c) SMLM images of the defect lines. d) AFM images of the defect lines. Two distinct groups of defects are produced, depending on the irradiation dose: hills for lower doses and holes for higher ones. e) Line profiles of the emitters in (c) with calculated average pitch distance. f) Line profiles of the AFM images in (d). Scale bars:
5mm (a), 1mm (c,d)}
\label{fig:si-line-patterns}
\end{figure}

\begin{figure}[ht]
\centering
\includegraphics[width=\linewidth]{figures/SI/fig-SXX-SBR calculation2.png}
\caption[Calculating Signal-To-Background Ratio values from processed widefield images]{Calculating Signal-To-Background Ratio values from processed widefield images (averaged stacks). Each emitter's position was manually selected, then the maximum intensity pixels were automatically found and averaged to get the signal value. To obtain the local background value the intensities of border pixels (indicated by red squares) were averaged. The final SBR value was achieved as the signal value divided by the background value. Pixel size: 100 nm }
\label{fig:si-sbr-calculation}
\end{figure}

\begin{figure}[ht]
\centering
\includegraphics[width=\linewidth]{figures/SI/fig-SXX-signal vs time in water.png}
\caption[SBR evolution in time]{a) Longer measurements of the SBR evolution from hole defects, indicating a possible saturation after 100 hours in water at room temperature. b) Alternative SBR metric: brightest pixel population (first decile) divided by the background (median), which doesn't require manual selection of emitters. A slight dip in SBR at the end might be due to defocus. }
\label{fig:si-sbr-alternative}
\end{figure}

\begin{figure}[ht]
\centering
\includegraphics[width=\linewidth]{figures/SI/fig-S3_fig3.png}
\caption[Additional characterization of optical properties of FIB-induced defects.]{Additional characterization of optical properties of FIB-induced defects. a)  Bleaching curves under 488 nm and 561 nm CW illumination. Double-exponent fit for the 488 illumination hints on two types of emitters being bleached. b) Fluorescence lifetime of hole- and hill-type defects under pulsed 561 nm illumination (background-corrected). While amplitude of fluorescence differs significantly, the lifetimes are quite similar, $\approx 1.1$ ns. c) Saturation curve for the hole-type defects (P$_{sat} \approx 50$\% of laser power, corresponding to $\approx 2.5$ kW$/cm^2$) d) ODMR curves for varying microwave power, for hole-type defects. Measurement done in air on a sample dried after a 30-hour immersion in water. }
\label{fig:si-optical-characterization}
\end{figure}

\begin{figure}[ht]
\centering
\includegraphics[width=\linewidth]{figures/SI/fig-SXX-afm-before-after-water.png}
\caption[AFM analysis of FIB-irradiated hBN flakes]{AFM analysis of FIB-irradiated hBN flakes before (a-c) and directly after (d-f) immersing the sample in water. The hole formation therefore happens on a timescale faster than a single high-resolution AFM scan ($\approx 20$ minutes). Scale bars: 1 $\mu m$ (a,d), 500 nm (b,e) }
\label{fig:si-afm-before-after-water}
\end{figure}

\begin{figure}[ht]
\centering
\includegraphics[width=\linewidth]{figures/SI/fig-SXX-afm-timelapse-in-water.png}
\caption[AFM timelapse of the same irradiated hBN flake immersed in water]{AFM timelapse of the same irradiated hBN flake immersed in water. a) After 22 hours. b) After 30 hours. c) Height profiles along respective lines in (a, b), showing that the holes are not changing their diameter ($\approx 400$ nm) much over time. Scale bars: 1 $\mu m$ (a,d), 500 nm (b,e) }
\label{fig:si-afm-timelapse-water}
\end{figure}

\begin{figure}[ht]
\centering
\includegraphics[width=\linewidth]{figures/SI/fig-SXX-afm-hole-size-vs-dose.png}
\caption[Dependence of hole size on irradiation dose (dwell time)]{Dependence of hole size on irradiation dose (dwell time). a) AFM images of isolated holes after water. b) Height profiles of their cross-sections. Scale bar: 100 nm }
\label{fig:si-afm-hole-size-vs-dose}
\end{figure}

\begin{figure}[ht]
\centering
\includegraphics[width=\linewidth]{figures/SI/fig-S66-AFM of TEM grids before water.png}
\caption[AFM analysis of suspended FIB-irradiated hBN flakes]{AFM analysis of suspended FIB-irradiated hBN flakes. a) Silicon nitride TEM grid with a thin hBN flake, irradiated with a 2D array of defects (pitch distance 600 nm, beam current 100 pA, dwell time 1.4 ms). b) Zoom-in into highlighted region in (a). c) Height profiles along respective lines in (b), showing that the flake is fully pierced by the ion beam. d) A thicker hBN flake on the same TEM grid, patterned with the same FIB parameters. e) Zoom-in into highlighted region in (d). f) Height profiles along respective lines in (e), showing that the flake is only partially milled by the ion beam. Scale bars: 1 $\mu m$ (a,d), 500 nm (b,e) }
\label{fig:si-tem-grids-afm}
\end{figure}

\begin{figure}[ht]
\centering
\includegraphics[width=\linewidth]{figures/SI/fig-SXX-AbsorptionBeforeAfterWater.pdf}
\caption[Absorption spectra of defects before and after water treatment]{Absorption spectra of defects before and after water treatment. The measurement was done in transmission using a white light source. The background measured next to the irradiated flake was substracted from the recorded signal. \textbf{@Aqeel please add/correct here}}
\label{fig:si-absorption}
\end{figure}

\begin{figure}[ht]
\centering
\includegraphics[width=\linewidth]{figures/SI/fig-S5_fig3.png}
\caption[Confocal spectral mapping of produced defects with varying irradiation dose]{Confocal spectral mapping of produced defects with varying irradiation dose. Emission pattern at a) 610 nm and b) 830 nm, corresponding to two spectral peaks observed in Fig. 3 in the main text. The emission at 830 nm seems to come from a larger ring around the defect site, while the 610 emission is more localized towards the center. Pixel size: 250 nm }
\label{fig:si-spectral-map}
\end{figure}

\begin{figure}[ht]
\centering
\includegraphics[width=\linewidth]{figures/SI/fig-SXX-SMLM montage_fig3.png}
\caption[Artefacts of SMLM analysis of defect sites]{a) Montage of a single FIB-induced defect site, emitting over 100 frames (2 seconds). Fluctuations in fluorescence intensity and spatial distributions are well visible, but the density of emitters is high, which prevents the localization of single emitters b) Same montage with an overlay of the SMLM detections, performed by ThunderSTORM\cite{ovesny2014thunderstorm} algorithm. One can notice multiple false localizations due to high emitter density, leading to artifacts in the reconstructed image below. Scale bar: 1 $\mu m$ }
\label{fig:si-smlm-montage}
\end{figure}

\begin{figure}[ht]
\centering
\includegraphics[width=\linewidth]{figures/SI/fig-SXX-WF-TS-SOFI.png}
\caption[Processed widefield images of fluorescent defects]{Processed widefield images of fluorescent defects. a) Averaged image of the stack of 10'000 frames with 20 ms exposure. b) Reconstructed image, summing up localizations of the same image stack, processed by the ThunderSTORM plugin\cite{ovesny2014thunderstorm}. c) Result of super-resolution optical fluctuation imaging (SOFI) processing\cite{Geissbuehler2012}, clearly displaying the grouping of emitters around the defect rim. Scale bar: 3 $\mu m$ (a-c), 500 nm (insets) }
\label{fig:si-wf-ts-sofi}
\end{figure}

\clearpage
\bibliography{FIBhBNtocite}